\documentclass[12pt]{article}
\usepackage[english]{babel}
\usepackage{amsfonts}
\usepackage[dvips]{graphicx}

\begin{document}

\title{\bf Pitot tube from Euler equations point of view. Application to Formula 1 cars.}

\date{September 2008}

\author{\bf Gianluca Argentini \\
\normalsize{[0,1]Bending - Italy}\\
\normalsize 01bending@gmail.com \\
\normalsize gianluca.argentini@gmail.com \\}

\maketitle

\begin{abstract}
The paper describes a mathematical model on the physical principles of the Pitot device for acquiring flows speed. The relation between speed and static pressures is derived from simple integration of the fluid dynamics equations for an inviscid flow. In this way, the treatment is general and is not based on the Bernoulli theorem on energy conservation, as usually done in scientific and technical literature. The application to Formula 1 cars is discussed.

\end{abstract}

\section{About Pitot tube concepts}

The Pitot tube (\cite{anderson}) is a mechanical device for measuring flow velocity by dynamic pressure. In its basic version, it has an L-shaped small channel, with the base-segment parallel to the flow, for capturing the value of total pressure $p_t = p + \frac{1}{2}\rho v^2$, and a vertical small channel, perpendicular to the flow direction, for acquisition of the static pressure value $p$ of the fluid flow where it is submerged (see Fig.1).\\
\noindent The French scientist Henry Pitot defined the basic concepts of the device in 1732 (\cite{levi}), and so before the formulation of Bernoulli theorem (1738) and Euler equations (1755) of inviscid fluid motion (see \cite{rouse}, pg. 48). In the usual explanations about the principles of Pitot's tube, the theorical results are derived from Bernoulli theorem on energy conservation, which is valid for every single streamlines of a flow tube (\cite{rouse}).\\
\noindent But using computational analysis one can see that the identification of a flow tube from the external domain to the internal of the Pitot L-channel is not clear, and the streamlines diverge near the frontal small hole of the device or they enter into the L-channel with vanishing speed (see Fig.2, Fig.3 and \cite{khan}, pg. 166).\\

\begin{figure}[h!]\label{PitotSchema}
	\begin{center}
	\includegraphics[width=6cm]{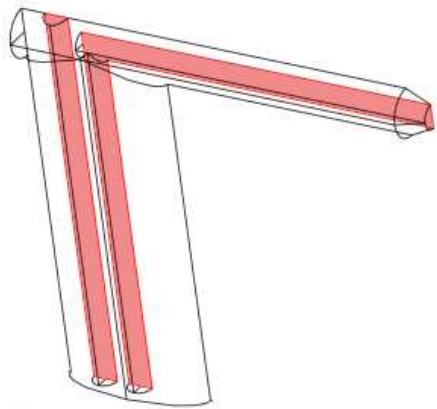}
	\caption{\it Pitot tube with its two channels.}
	\end{center}
\end{figure}

\begin{figure}[ht]\label{streamlines}
	\begin{center}
	\includegraphics[width=10cm]{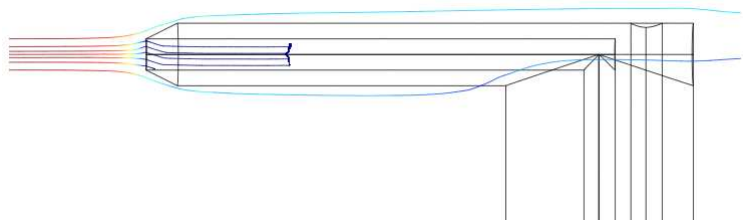}
	\caption{\it Computed streamlines near the front of the Pitot tube.}
	\end{center}
\end{figure}

We would show that the formula, used by Pitot, for estimating the flow speed can be obtained from a point of view more general than the Bernoulli theorem. We use the Navier-Stokes equations in the case of an inviscid fluid flow, that is the Euler equations.

\begin{figure}[h!]\label{flusso3D}
	\begin{center}
	\includegraphics[width=10cm]{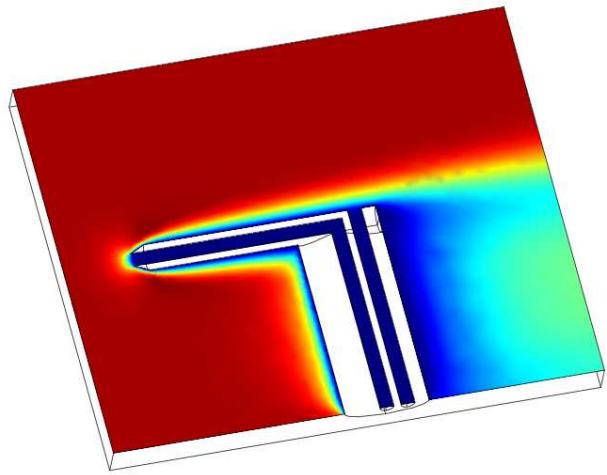}
	\caption{\it Speed field on the central surface of the pitot, from computational simulation.}
	\end{center}
\end{figure}

\section{Using Euler equations}

We need the following equations for inviscid steady flow, refered to a cartesian system $\{x,y\}$, where $p$ is the static pressure, $\rho$ the fluid density and ${\bf u}=(u_1, u_2)$ is the velocity field:

\begin{eqnarray}\label{euler}
	\nonumber \rho \left( u_1\frac{\partial u_1}{\partial x} + u_2\frac{\partial u_1}{\partial y} \right) = - \frac{\partial p}{\partial x}\\
	\rho \left( u_1\frac{\partial u_2}{\partial x} + u_2\frac{\partial u_2}{\partial y} \right) = - \frac{\partial p}{\partial y}
\end{eqnarray}

\noindent We use the hypothesis that the density $\rho$ is constant and that the speed of fluid, inside the channels of the Pitot tube, is zero, which is a good approximation.\\
\noindent At first, consider the L-shaped channel. If the $x$-axes is defined to be parallel to the channel, $u_2=0$ and the flow becomes one-dimensional:

\begin{equation}\label{eulerOneD}
	\rho u_1\frac{\partial u_1}{\partial x} = - \frac{\partial p}{\partial x}\\
\end{equation}

\noindent Let $v$ and $p_e$ the constant speed and the static pressure of the fluid in the domain external to the Pitot channel, and $p_i$ the static pressure in the interior. Integrating previous equation from an external point $e$ to an internal point $i$, we have

\begin{eqnarray}\label{eulerOneDintegrated}
	\nonumber \rho \int_e^i u_1\frac{\partial u_1}{\partial x} dx = \frac{1}{2}\rho \left[u_{1,i}^2 - u_{1,e}^2 \right] =\\
	= - \frac{1}{2}\rho v^2 = - \left( p_i - p_e \right)
\end{eqnarray}

\noindent so we obtain

\begin{equation}\label{bernoulli}
	p_i = p_e + \frac{1}{2}\rho v^2
\end{equation}

\noindent This formula, which is the expression of the Bernoulli theorem, gives the information that the pressure in the L-shaped Pitot channel is the sum of static and dynamic pressure of external flow.\\

Let now consider the vertical channel. Computational simulations show that the $y$-component $u_2$ of the velocity field is almost everywhere null in the region near the hole of this channel, but some significative variations in gradient $\nabla u_2$ can occur. In this case, even if we assume $u_2=0$, the second Euler equation is

\begin{equation}
	\rho u_1\frac{\partial u_2}{\partial x} = - \frac{\partial p}{\partial y}
\end{equation}

\noindent and a simple integration is not possible as in previous case. For avoiding this problem, we use a change of cartesian coordinates system.

\section{Changing cartesian reference}

Let be $\{X,Y\}$ the new cartesian system by clockwise rotation of angle $\alpha$ of the old $\{x,y\}$.

\begin{figure}[h!]\label{reference}
	\begin{center}
	\includegraphics[width=8cm]{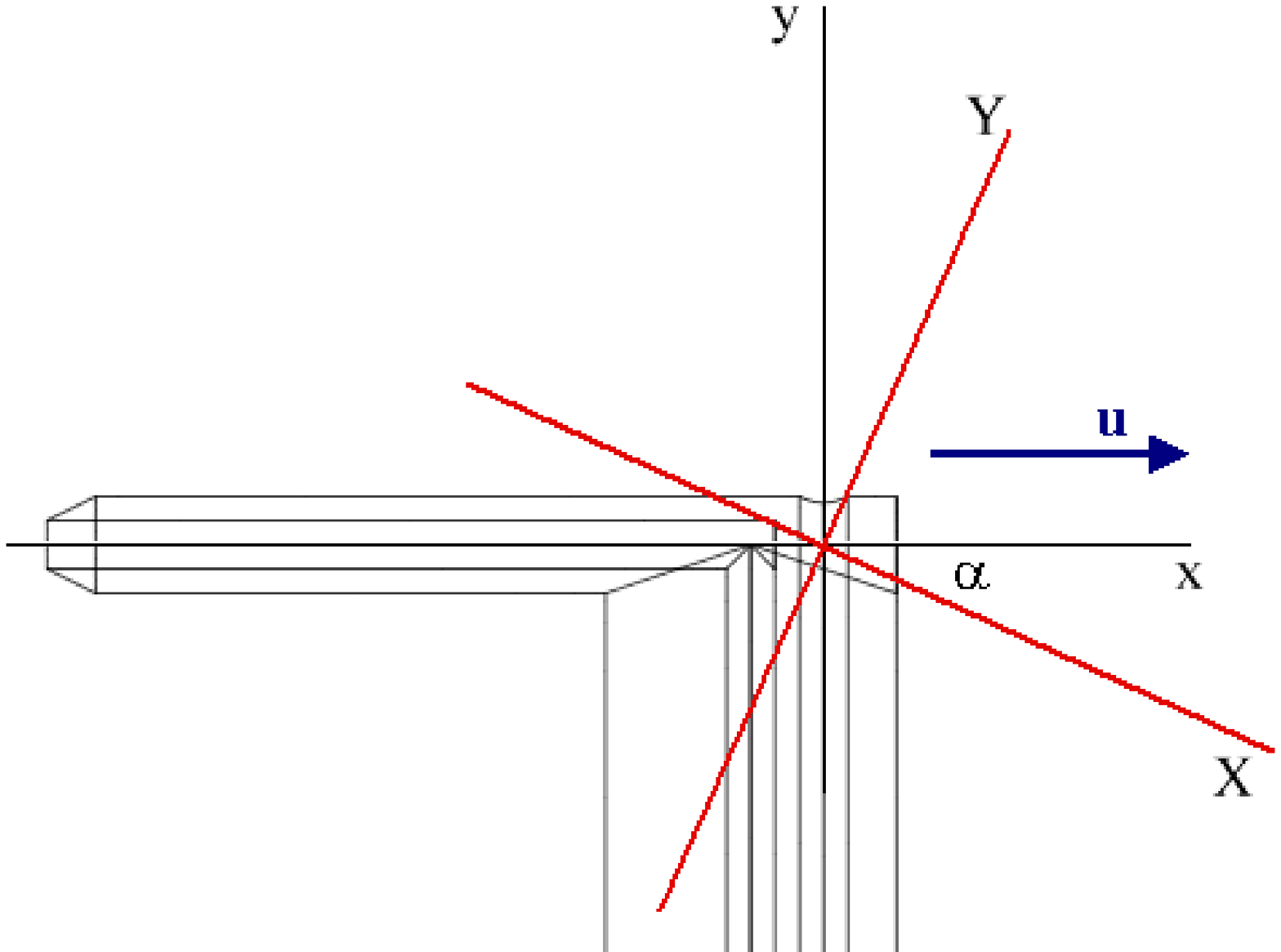}
	\end{center}
\end{figure}

\noindent The transformation rules of this change of coordinates are

\begin{equation}\label{transformationRules}
\left\{
\begin{array}{ll}
	\nonumber x = X cos\alpha - Y sin\alpha\\
	y = X sin\alpha + Y cos\alpha
\end{array}
\right.
\end{equation}

\noindent The velocity vector $\bf u$ in the new coordinates becomes $\bf U$ = $\left(U_1,U_2\right)=\left( v cos\alpha, - v sin\alpha \right)$, and if $q=q(X,Y)=p(x(X,Y),y(X,Y))$ is the static pressure, the Euler equations in the $\{X,Y\}$ system have the form ($\rho$ is a constant)

\begin{eqnarray}\label{eulerXY}
	\nonumber \rho \left( U_1\frac{\partial U_1}{\partial X} + U_2\frac{\partial U_1}{\partial Y} \right) = - \frac{\partial q}{\partial X}\\
	\rho \left( U_1\frac{\partial U_2}{\partial X} + U_2\frac{\partial U_2}{\partial Y} \right) = - \frac{\partial q}{\partial Y}
\end{eqnarray}

\noindent Using the fact that $v$ is, from hypothesis, a function only of the $x$ variable, from chain derivation rule (see \cite{courant}) we have the following identities:

\begin{eqnarray}\label{identities}
	\partial_X U_1 = \partial_x (vcos\alpha)\partial_X x = cos\alpha \hspace{0.1cm}\partial_x v \hspace{0.1cm} cos\alpha = cos^2\alpha \hspace{0.1cm} \partial_xv\\
	\partial_Y U_1 = \partial_x (vcos\alpha)\partial_Y x = cos\alpha \hspace{0.1cm}\partial_x v \hspace{0.1cm} \left(-sin\alpha \right) = -sin\alpha \hspace{0.1cm} cos\alpha \hspace{0.1cm} \partial_xv\\
	\partial_X U_2 = \partial_x (vsin\alpha)\partial_X x = sin\alpha \hspace{0.1cm}\partial_x v \hspace{0.1cm} cos\alpha = sin\alpha \hspace{0.1cm} cos\alpha \hspace{0.1cm} \partial_xv\\
		\partial_Y U_2 = \partial_x (vsin\alpha)\partial_Y x = sin\alpha \hspace{0.1cm}\partial_x v \hspace{0.1cm} \left(-sin\alpha \right) = - sin^2\alpha \hspace{0.1cm} \partial_xv\\
	\partial_X q = \partial_x p \partial_X x + \partial_y p \partial_X y = cos\alpha \hspace{0.1cm} \partial_x p + sin\alpha \hspace{0.1cm} \partial_y p\\
	\partial_Y q = \partial_x p \partial_Y x + \partial_y p \partial_Y y = -sin\alpha \hspace{0.1cm} \partial_x p + cos\alpha \hspace{0.1cm} \partial_y p
\end{eqnarray}

\section{The final Pitot formula}

Now, for computing the pressure in the vertical channel, consider $\alpha = -\frac{\pi}{2}$. In this case, using previous transformation rules, we note that $U_1 = 0$, $\partial_Y U_1 = 0$ and $\partial_X q = -\partial_y p$, so that the first Euler equation in (\ref{eulerXY}) becomes

\begin{equation}
	\partial_y p = 0
\end{equation}

\noindent therefore, by integration from an external point $e$ to an internal point $i$, we have $p_i = p_e$: the pressure in the Pitot channel perpendicular to the flow is the same as the static pressure of the external flow.\\

We can now find the formula for estimating the flow speed $v$ using the two informations obtained by the Pitot tube. Let $A$ the value of $p_i$ registered by the L-shaped channel and $B$ the value registered by the other channel. We have shown that 

\begin{eqnarray}\label{PitotSystem}
	\nonumber  A = p_e + \frac{1}{2}\rho v^2\\
	B = p_e
\end{eqnarray}

\noindent Solving the system for $v$, and noting that $A = p_t$, the final formula is

\begin{equation}\label{PitotSpeed}
	v = \sqrt{\frac{2(p_t - p_e)}{\rho}}
\end{equation}

\noindent which is the usual Pitot expression of the flow speed (see \cite{anderson}, \cite{khan}).

\section{Single channel Pitot tubes}

In the case where there is no need of a realtime flow speed acquisition, but only of defered measurements (e.g. values of flow rate in industrial devices as pumps or valves), it can be useful a more simple Pitot tube having only the L-shaped channel. With the tube geometrically placed opposite to the flow direction, one have the value of the total pressure $p_t$. With the tube rotated of an angle $\pi$ around the horizontal direction perpendicular to the flow, one have a second value $p_i$. What is the meaning of this $p_i$?\\
\noindent Suppose the origin of the cartesian reference $\{x,y\}$ is on the frontal hole of the channel. From the tranformation laws, we have $x=-X$, $y=-Y$, $U_1=-v$, $U_2=0$, $\partial_X q = -\partial_x p$. In the rotated new cartesian system $\{X,Y\}$ the first Euler equation becomes

\begin{equation}
	\rho U_1\frac{\partial U_1}{\partial X} = - \frac{\partial q}{\partial X}
\end{equation}

\noindent By integration of the first member, on the variable $X$, from a point $+A$ (that is, internal to the tube) to a point $-A$ (that is, external to the tube), and using the (good) approximation that there is no flow inside the channel, we have

\begin{equation}
	\int_{+A}^{-A}\rho U_1\frac{\partial U_1}{\partial X} dX = \frac{1}{2}\rho \left[U_1^2(-A)-U_1^2(+A)\right] = \frac{1}{2}\rho U_1^2(-A) = \frac{1}{2}\rho v^2
\end{equation}

\noindent For the second member, recalling that $dX = -dx$ and using the rules of integration by change of variable (see \cite{courant}), we have

\begin{eqnarray}
	\nonumber \int_{+A}^{-A}- \frac{\partial q}{\partial X} dX = \int_{-A}^{+A}\frac{\partial p}{\partial x}(-dx) =\\
	\int_{+A}^{-A} \frac{\partial p}{\partial x} dx = p(-A)-p(+A) = p_e - p_i
\end{eqnarray}

\noindent Comparing the two previous results, note that

\begin{equation}
	p_i = p_e - \frac{1}{2}\rho v^2
\end{equation}

\noindent that is the $\pi$-rotated Pitot tube gives the value of the difference between the flow static and dynamic pressure.\\

\noindent Now, suppose we have these two measures $m_0$ and $m_{\pi}$ by this single channel Pitot tube. From the system

\begin{eqnarray}\label{PitotSystem2}
	\nonumber  m_0 = p_e + \frac{1}{2}\rho v^2\\
	m_{\pi} = p_e - \frac{1}{2}\rho v^2
\end{eqnarray}

\noindent we can estimate the flow speed:

\begin{equation}\label{PitotSpeed2}
	v = \sqrt{\frac{2(m_0 - m_{\pi})}{\rho}}
\end{equation}

\section{Pitot tube and Formula 1 cars}

In the last years the formulas 1 have small Pitot tubes on the front nose for a realtime acquisition of the car speed on the track. Usually, for acquiring the value of the flow static pressure $p_e$, these tubes have more than one hole placed on an ideal circumference around the external wall of the first segment of the L-shaped channel.

\begin{figure}[h!]\label{formula1Pitot}
	\begin{center}
	\includegraphics[width=10cm]{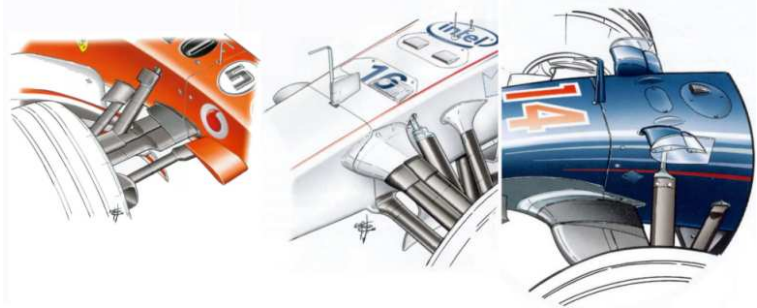}
	\caption{{\it Pitot tubes on the front nose of some Formula 1 cars. Graphics by} \normalsize{Giorgio Piola (\cite{piola})}.}
	\end{center}
\end{figure}

\noindent In the 2007 season, some teams (as BMW-Sauber, Ferrari, Toyota, Williams) have used high tubes with the upwind segment of the L-shaped channel placed parallel to the track, that is parallel to the direction of the freestream. Some other teams (as McLaren, Red Bull, Renault, Toro Rosso) instead have used short tubes with the upwind segment parallel to the nose profile, that is parallel to the direction of the flow on the boundary of the car shape. In the first case, the car speed $v_{\infty}$ can be directly computed by (\ref{PitotSpeed}). In the second case, the device gives the speed of the flow along the nose profile. Let be $\{x,y\}$ a cartesian system with $x$-axes parallel to the track and $y = f(x)$ the analytical expression of the nose-profile. If $x_0$ is the $x$-position of the Pitot tube, the device gives a speed $v_P$ related to $v_{\infty}$ by the relation

\begin{eqnarray}\label{vAngleTemp}
	\nonumber v_P = v_{\infty} \hspace{0.1cm} cos\alpha = v_{\infty} \hspace{0.1cm} cos \left(arctan f'(x_0) \right) =\\
	= v_{\infty}\frac{1}{\sqrt{1+f'^2(x_0)}}
\end{eqnarray}

\begin{figure}[h!]
	\begin{center}
	\includegraphics[width=8cm]{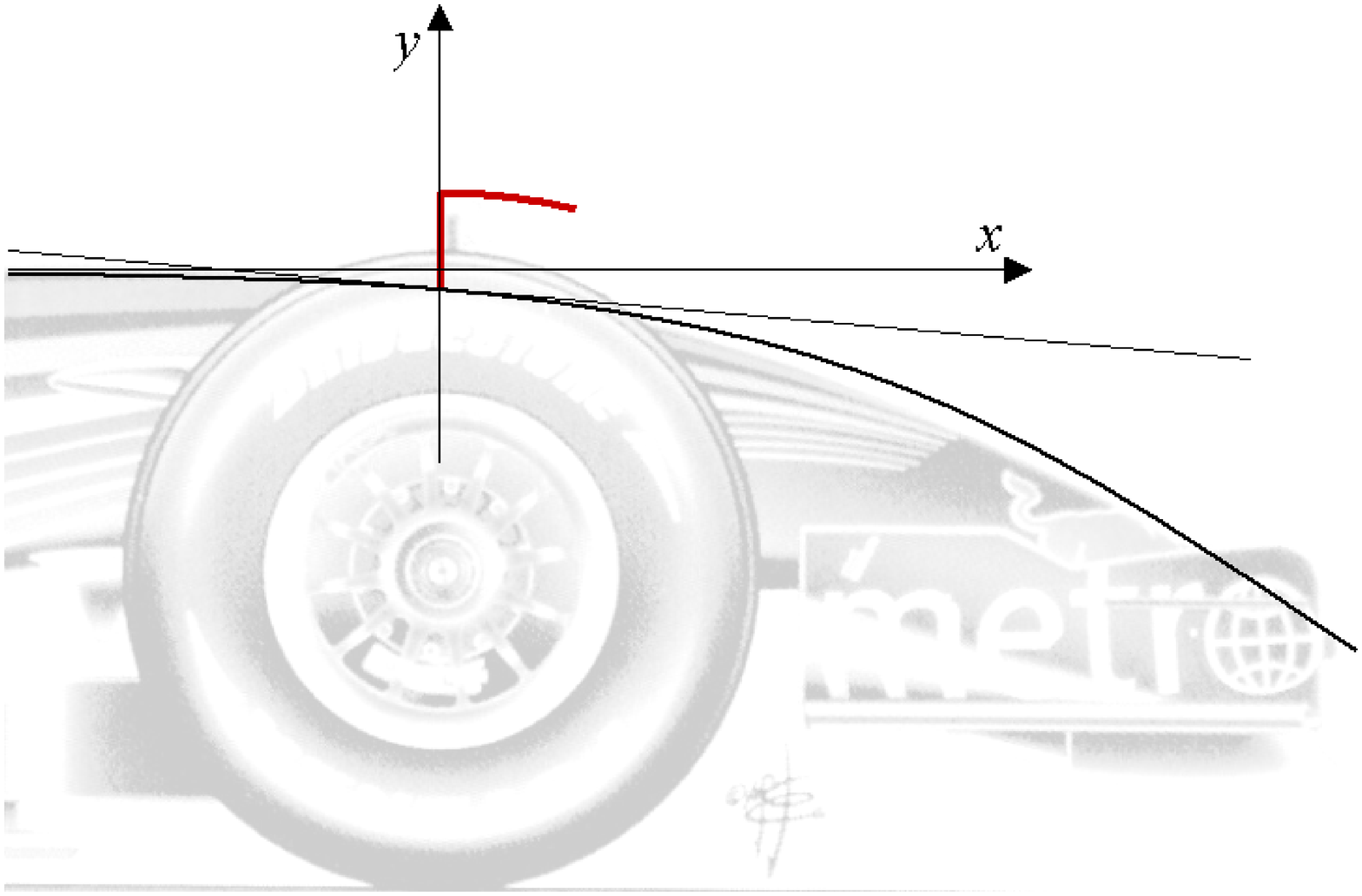}
	\end{center}
\end{figure}

\noindent Usually the geometrical position of the pitot on the nose is so that $\alpha$ is a small angle, therefore we can consider the following Taylor series expansions:

\begin{eqnarray}
	\nonumber f'(x_0) = tan\left(\alpha\right) \approx \alpha\\
	\sqrt{1+t^2} \approx 1 + \frac{t^2}{2}
\end{eqnarray}

\noindent Then, from (\ref{vAngleTemp}), the freestream speed, that is the speed car on the track, can be express by

\begin{equation}\label{vCar}
	v_{\infty} = \frac{2+\alpha^2}{2}\hspace{0.1cm}v_P
\end{equation}

\noindent Note that, in this configuration, if one doesn't consider tha value of the angle $\alpha$, even if very small, the car speed given by the Pitot device can be a little small than the real speed. For example, if $\alpha=$ 0.2 $rad$ and $v_P$ is 300 $km/h$, the real car speed computed by (\ref{vCar}) is $v_{\infty}$ = 306 $km/h$. This is a significative difference to be considered in a world as that of Formula 1.

\begin{figure}[h!]
	\includegraphics[width=2cm]{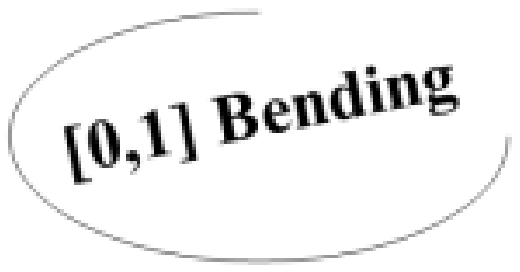}
\end{figure}
\noindent \tiny {\bf Gianluca Argentini}, mathematician, works on the field of\\
fluid dynamics and optimization of shapes for bodies\\
moving inside fluid flows. He has found {\bf [0,1]Bending},\\
a Design Studio in Italy for scientific and industrial applications.


\begin{thebibliography}{9}

\bibitem{anderson} J.D.Anderson, {\it Fundamentals of Aerodynamics}, McGraw-Hill, 1984

\bibitem{courant} R.Courant, {\textit{Differential and Integral Calculus}}, Vol. 1 and 2, Wiley, New York, 1988

\bibitem{khan} I.A.Khan, {\it Fluid Mechanics}, Oxford University Press US, 1995

\bibitem{levi} E.Levi, D.Medina, {\it The Science of Water: the Foundation of Modern Hydraulics}, ASCE Publications, 1995

\bibitem{piola} G.Piola, {\it Formula 1 2006-2007 Technical Analysis}, Giorgio Nada Editore, 2007

\bibitem{rouse} H.Rouse, {\it Fluid Mechanics for Hydraulic Engineers}, Dover, New York, 1961

\end{thebibliography}
\end{document}